# Spatial Correlation in Matter Wave Interference as a Measure of Decoherence, Dephasing and Entropy

Zilin Chen, Peter J. Beierle, and Herman Batelaan

*Department of Physics and Astronomy, University of Nebraska-Lincoln – 208 Jorgensen Hall, Lincoln, NE 68588-0299, USA*

**Abstract –** The loss of contrast in double-slit electron-diffraction due to dephasing and decoherence processes is studied. It is shown that the spatial correlation function of diffraction patterns can be used to distinguish between dephasing and decoherence. This establishes a measure of time-reversibility that does not require the determination of coherence terms of the density matrix, while von Neumann entropy, another measure of time-reversibility, does require coherence terms. This technique is exciting in view of the need to understand and control the detrimental experimental effects of contrast loss and for fundamental studies on the transition from the classical to the quantum regime.

Loss of contrast can be attributed to physical processes divided into two broad classes. Dephasing processes are time-reversible, while decoherence processes are time-irreversible. Time-reversibility can be established by evaluating the change in entropy, $S = -Tr\rho \ln \rho$, where $\rho$ is the density matrix describing the physical system. When $S$ remains constant in time, the process is time-reversible; while when it increases in time, the process is time-irreversible [1]. The value of the entropy depends on the off-diagonal or coherence terms of the density matrix, which is apparent from the calculation of the entropy using the spectral decomposition, $S = -\sum \lambda_i \ln \lambda_i$, where $\lambda_i$ are the eigenvalues of the density matrix. In diffraction experiments, determination of the coherence terms would require special techniques, such as quantum state tomography [2]. Thus, the on-diagonal terms of the density matrix, that describes the spatial probability distribution of the physical system, do not appear to provide direct access to the very nature of the process that is limiting the contrast observed in that probability distribution. This makes it hard to identify sources of contrast loss and thus take appropriate measures to reduce such a loss. Additionally, when studying the transition from the quantum to classical domain, by introducing controlled decoherence processes, it is hard to establish that it is indeed decoherence and not dephasing that causes a loss of contrast.

In this paper, we propose and analyze a method based on repetitive measurements of the spatial probability distributions, that can be used to distinguish dephasing from decoherence processes. The spatial correlation function of the measurements provides this information. For dephasing processes that upon visual inspection appear to completely destroy the diffraction pattern, the correlation function restores the far-field diffraction pattern. For decoherence processes no such restoring works.

To support our claims, we consider an electron double-slit experiment [3,4] as an archetypical example of an interference experiment, and add a process by which contrast is lost. This situation described is not just a thought experiment, but is typical for real experiments. For example, we reported an electron diffraction experiment with nano-fabricated gratings, where some loss of contrast was observed and modeled [4].

An optical experiment that exhibited loss of contrast was performed by Rui-Feng et al. [5]. In their setup, a laser beam with a 632.8 nm wavelength passed through a ground glass disk and double-slit. The detection screen was placed in the Fresnel diffraction region with respect to the double-slit. The ground glass disk appeared to completely destroy the contrast of the diffraction pattern. The normalized second order correlation function was used to regain the double-slit diffraction pattern. This is a striking result in its own right. The central question which was not addressed is: "Does the ground glass disk dephase or decohere the laser light?" Or in general: "Can spatial correlation be used to identify dephasing and decoherence processes?"

In our simulation, we studied the analogous double-slit physical system but changed the diffracting particle from photons to electrons as decoherence theory is often studied in the context of matter optics [6–11]. Based on the matter-wave analogy [12] the method is expected to work for both matter waves and optics. Our approach is to simulate electron diffraction in three different situations as shown in Fig. 1. In the first situation, a two-path interferometer, i.e., a double-slit experiment, exhibits excellent contrast. In the second and third situation, an object is introduced after the double-slit that interacts with the electron wave so to cause dephasing or decoherence. We will refer to this as a "dephaser" and "decoherer." These latter two patterns share a reduced contrast but are found to have qualitatively different correlation functions.

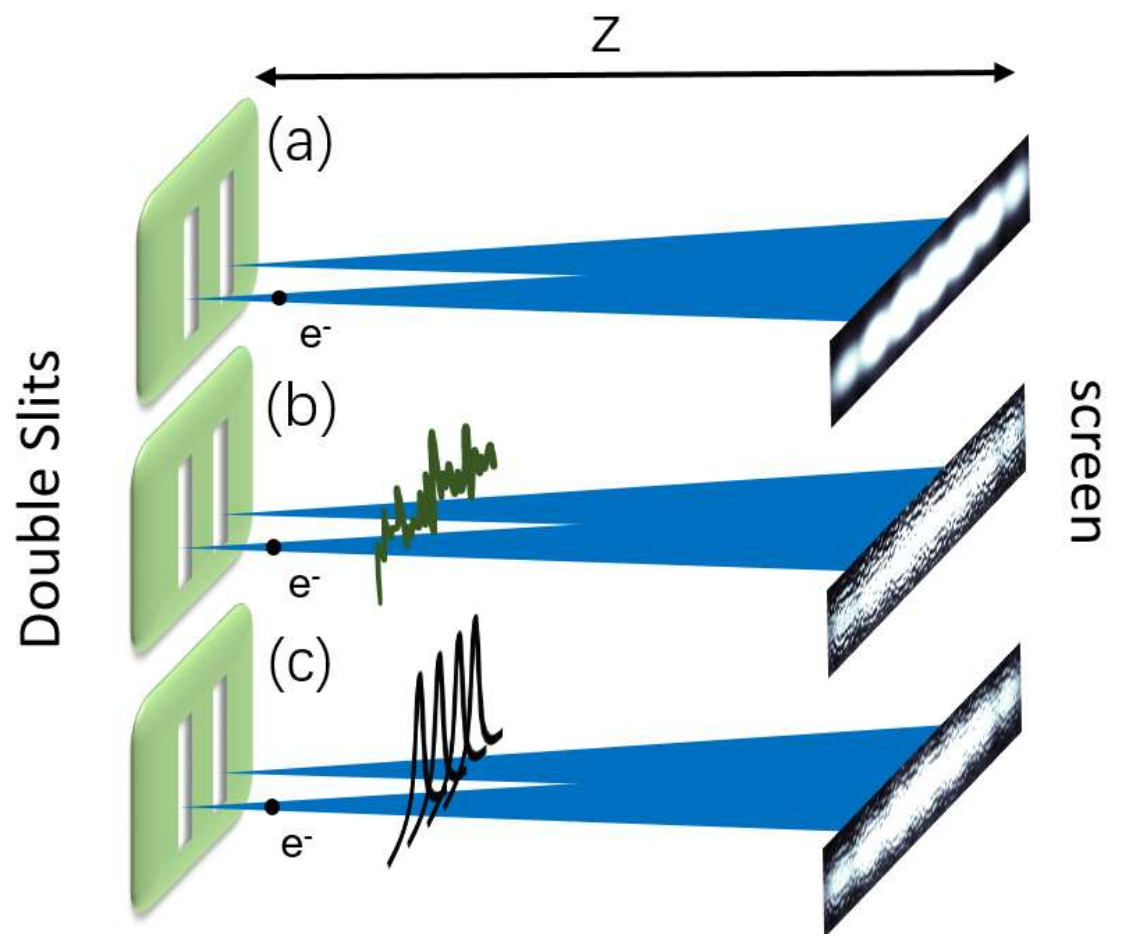

FIG. 1. Dephasing versus decoherence. A sketch of three situations where electron waves interfere. In (a), electron waves are unaffected. An interference pattern will be observed in the probability distribution with excellent contrast. The observed pattern is taken from Bach et al. [4]. In (b), a "dephaser," represented by a random potential distribution (green wiggly line), adds a position dependent phase to the electron wave. As a result, the diffraction pattern appears to be washed out, but information about the diffraction pattern can be recovered. In (c), a "decoherer" separates the electron wave into a probabilistic sum of several Gaussian waves in addition to a random potential distribution. A low contrast diffraction pattern similar to (b) is obtained, but now the pattern cannot be recovered.

The example physical system we study (Fig. 2) is motivated by previously used experimental parameters for electron double-slit and decoherence experiments [10,13]. A coherent electron wave with an energy of $E = 1670$ eV and a transverse width of $w_0 = 15$ μm at the source propagates $L_1 = 24$ cm and encounters a double-slit. A dephaser or decoherer is located immediately after the double-slit, represented by a horizontal surface. The size of the surface and the double-slit are both chosen to be 500 nm wide, larger than the distance between the center of the two slits ($D =$ 150 nm). The width of the slits is $d = 50$ nm. Electrons are diffracted and dephased or decohered and continue to propagate to the detection screen at a distance $L_2 = 25$ cm. An interference pattern can be found on the detection screen which is placed in the far-field region (or in the Fresnel diffraction region with respect to the double-slits but far-field region with respect the single slit (as in [5]).

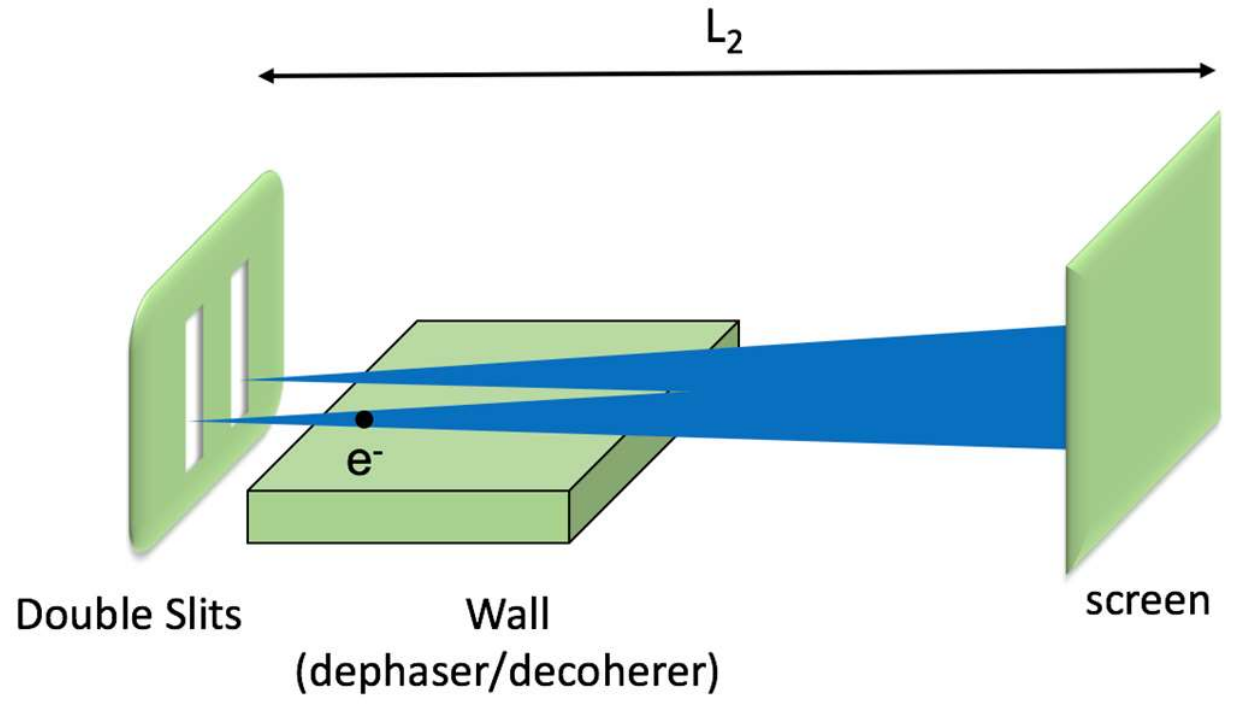

FIG. 2. Schematic of the physical system. An electron wave impinges on a double slit. Subsequently, a dephaser or decoherer disturbs the electron wave. The corresponding real phenomenon is plausibly caused by the back-action of the image charge on the electron [10,14–18]

Propagation of the electron wave is simulated using the path integral method [12,19],

$$\psi_k\left(x_k\right) = \int_{-w/2}^{w/2} dx' \psi_1\left(x'\right) h_k\left(x_k, x'\right). \tag{1}$$

The state of the wave-function at each location $x$ at the double-slit plane has accumulated a phase attribution from the state of the wave function at each location $x'$ according to the impulse response function $h_k\left(x_k, x'\right)$. Subsequently, the wave-function is modified by the dephaser or decoherer and finally propagated from the double-slit to the detection screen. The impulse response function [20],

$$h_k\left(x, x'\right) = e^{i2\pi l/\lambda} T\left(x'\right), \tag{2}$$

is given in terms of the transmission function

$$T\left(x'\right)=e^{i\theta\left(x'\right)}D\left(x'\right), \tag{3}$$

where $l=\sqrt{(x-x')^2+z^2}$ is the propagation length, $z$ is either $L_1$ or $L_2$, and $\lambda = h/\sqrt{2mE}$ is the de Broglie wavelength of electron. The double-slit transmission function $D\left(x'\right)$ equals one at the slits, and zero elsewhere.

To describe the dephaser, the random phase $\theta\left(x'\right)$ is given by a sum of Gaussians,

$$\theta\left(x'\right)=\sum_i A_i e^{-\left(\frac{x'-x_i}{\sqrt{2}\sigma_i}\right)^2}, \tag{4}$$

where $A_i$ are uniformly distributed random numbers ranging from 0 to 2π. The Gaussian widths $\sigma_i$ are random numbers with a normal distribution. The mean value of $\sigma_i$ is chosen at 4 nm, and its standard deviation at 1 nm for our numerical example. The set of coordinates of centers $\{x_i\}$ are uniformly distributed random positions covering the double slit. Thus, to realize this dephaser, 500 different Gaussian distributions are combined [13]. The spacing of the random Gaussians is much smaller than slit width (50 nm) and results in a probability distribution that is spread all over the detection screen.

To describe the decoherer, the wave front is cut into $N$ independent overlapping Gaussians, $\varphi_n\left(x'\right)$, effectively reducing the tranverse coherence length, $w$, to the FWHM of the Gaussian, $2\sqrt{2\ln 2}\sigma_i$. The Gaussians are propagated separately to find the wave-functions $f_n\left(x'\right)$ at the detection screen location. The density matrix before the decoherer is given by

$$\rho_i\left(x,x'\right)=\sum_{n=1,N}\frac{1}{N}\varphi_n\left(x\right)\varphi_n^*\left(x'\right), \tag{5}$$

with

$$\varphi_n\left(x'\right)=\sqrt{N}\psi\left(x'\right)e^{i\theta\left(x'\right)}e^{-\left(\frac{x'-x_n}{\sqrt{2}\delta_0}\right)^2}. \tag{6}$$

The normalization constant $N$ is the same for each Gaussian, $\psi(x')$ is the initial wavefront, and $\delta_0$ is a constant width of 100 nm. Each Gaussian is shifted by $x_n = nx_0$, with $x_0 = 12.5$ nm. Additionally, the same smooth potential phase shift $e^{i\theta(x')}$ described in Eqs. 3 and 4 is applied. The final density matrix is given by

$$\rho_f\left(x,x'\right)=\sum_{n=1,N}\frac{1}{N}f_n\left(x\right)f_n^*\left(x'\right). \tag{7}$$

The von Neumann entropy is calculated from $S=-\sum \lambda_i \ln \lambda_i$, where $\lambda_i$ are the eigenvalues of the density matrix. The initial entropy for the pure state is $S=0$. The effect of the decoherer is to reduce the absolute value of the off-diagonal matrix elements in the density matrix. Consequently, for the dephaser the entropy remains the same, while for the decoherer the entropy increases (Fig. 3).

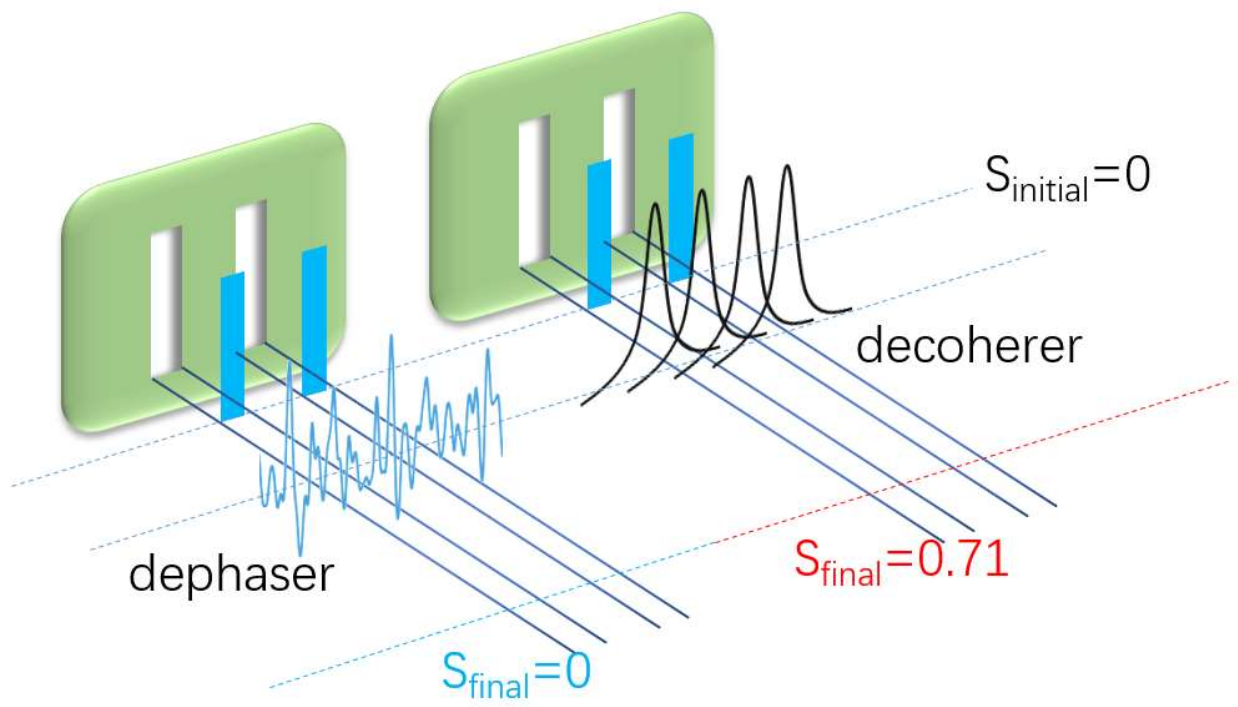

FIG. 3. Entropy. Schematic representation of the decoherer and dephaser in the simulation. The blue vertical lines show the electron wave front after passage through the double-slit. The Gaussians represent a sample of four incoherent waves emerging from the decoherer. The blue curve is a sample of the smooth random potential added to the electron wave for both the dephaser and decoherer. The entropy before and after the dephaser/decoherer is indicated.

The dependence of the von Neumann entropy on the transverse coherence length, $w$, after the decoherer, is determined, and compared to the Shannon entropy. To do this, the Gaussians widths and centers were varied, keeping their ratio fixed. A narrower width describes more decoherence and yields larger entropy and vice versa. A simpler decoherer model is added for comparison. In this model, the overlapping Gaussian functions are replaced with adjacent non-overlapping top-hat functions. The Shannon entropy, $S=-\sum p_i \ln p_i$, for this simpler decoherer, can be calculated analytically, and compared to the computer simulated von Neumann entropy. Here, $p_i$, is the probability to land within one top-hat function. When the two slits are covered with N top-hat functions, the probability for an electron to be found within one of the top-hats, is $p_i = w/2d$. The corresponding Shannon entropy is $S=-\sum p_i \ln p_i = \ln(2d/w)$. The analytic Shannon entropy matches the simulated von Neumann entropy very well (Fig. 4). This agreement indicates that our decoherer behaves as expected. For the case of Gaussian distributions, the entropy matches that for the top-hat case very well when the widths, $w$, are smaller than a single slit. As the value of $w$ is increased above the single slit width, but remains below the slit separation, the entropy remains relatively constant. When $w$ starts to exceed the separation, the entropy reduces to zero, as expected for a fully coherent state.

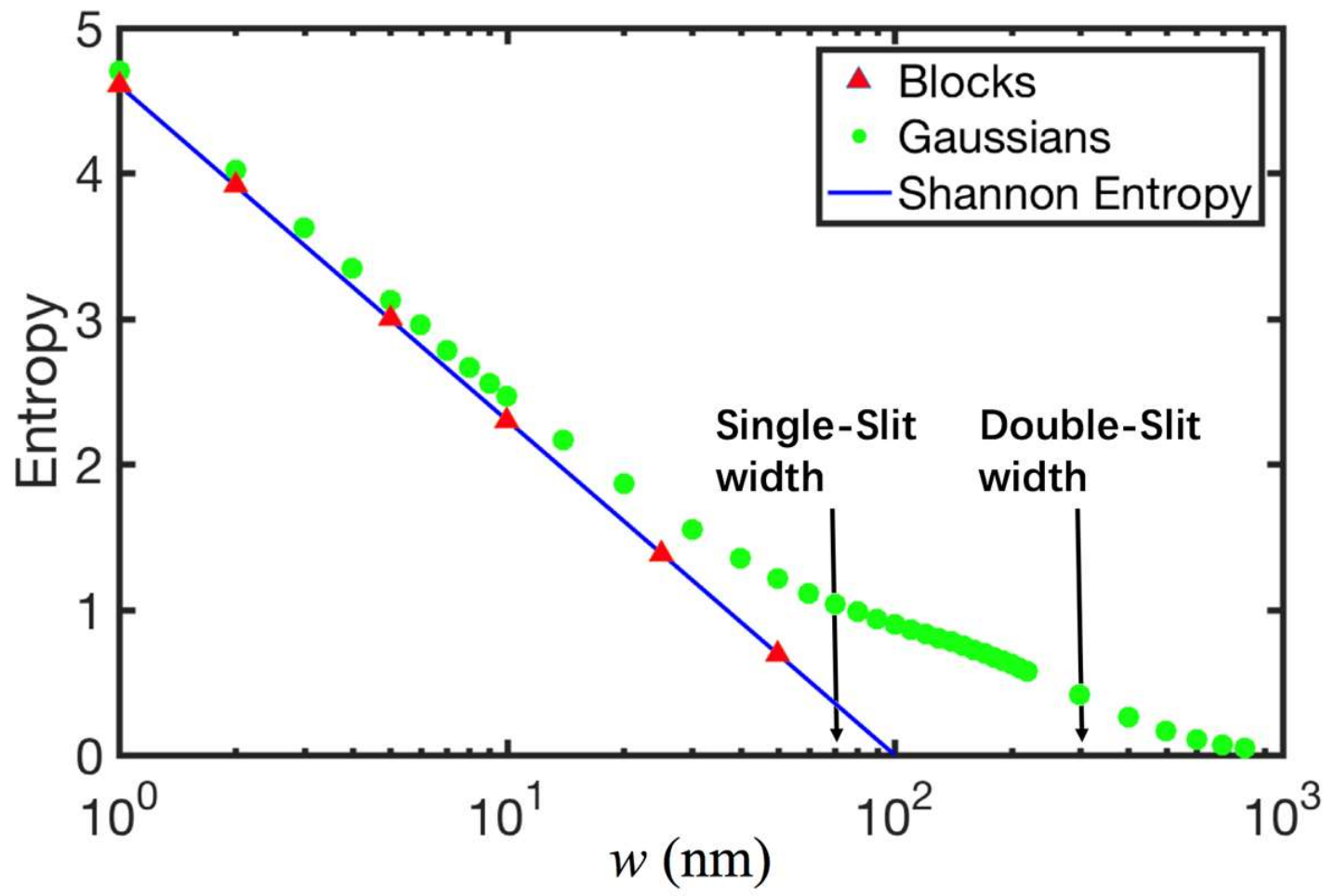

FIG. 4. Transition of Entropy. The entropy change is indicated as a function of the transverse coherence length. The von Neumann entropy for the Gaussian model (green circles), and the top-hat model (red triangles) are compared to the Shannon entropy (blue line). The von Neumann entropy and Shannon entropy (blue line) match well for the top-hat model. The Gaussian model matches well for transverse coherence lengths smaller then a single-slit. When the transverse coherence length exceeds the slit separation, the entropy approaches zero as expected for a pure state.

Now that we have introduced a dephaser and decoherer, we can proceed to test if a repetitive measurement of the probability distribution can be used to independently determine if a process is due to dephasing or decoherence. To do so, the diffraction pattern was calculated 500 times for both the dephaser and decoherer. Each realization used a different set of random numbers to generate a dephaser and decoherer. In Fig. 5, two realizations are shown. In the dephasing realizations 5(a) and 5(c), the peaks and valleys are more pronounced than in the decoherer realizations 5(b) and 5(d). This is consistent with earlier work using a Wigner function approach [21]. Panels (e) and (f) are averaged patterns over 500 realizations.

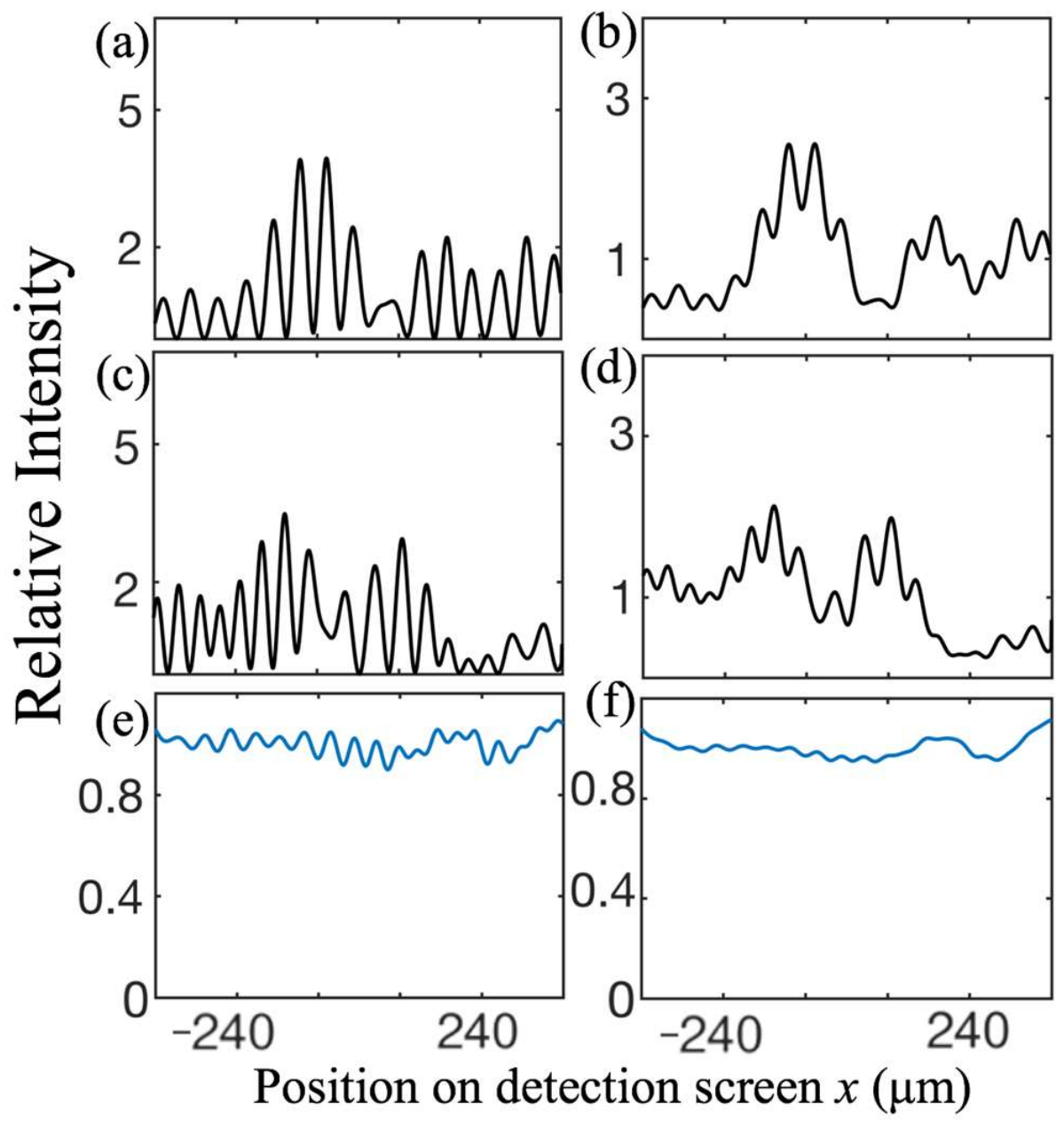

FIG. 5. Simulation realizations. (a) and (c) are two realizations with a dephaser and (b) and (d) are with decoherer. In (a), (b) and (c), (d), the same random number are used thus the influence of dephaser and decoherer will be more comparable. (e) and (f) are dephaser/decoherer averaged pattern over 500 realizations. The decoherer gives a similar probability distribution to dephaser but with blurred peaks.

For each probability distribution, the correlation function is calculated. The second order intensity correlation function is defined as,

$$G^{(2)}\left(x_1', x_2', x_1'', x_2''\right) \equiv \left\langle \psi_1^*\left(x_1'\right)\psi_2^*\left(x_2'\right)\psi_1\left(x_1''\right)\psi_2\left(x_2''\right)\right\rangle. \tag{8}$$

from which it follows that $G^{(2)}\left(x_1, x_2\right) = \left\langle I_1\left(x_1\right) I_2\left(x_2\right)\right\rangle$ . Where $x_1$ and $x_2$ are coordinates of probability distribution. Following Cheng's derivation [20]

$$\begin{aligned} &G^{(2)}\left(x_1, x_2\right) \\ &= \int dx_1' dx_2' dx_1'' dx_2'' G^{(2)}\left(x_1', x_2', x_1'', x_2''\right) \\ &\times h_1\left(x_1, x_1'\right) h_2\left(x_2, x_2'\right) h_1^*\left(x_1, x_1''\right) h_2^*\left(x_2, x_2''\right). \end{aligned} \tag{9}$$

Applying the result from Goodman [22]

$\left\langle u_1^* u_2^* u_3 u_4\right\rangle \equiv \left\langle u_1^* u_3\right\rangle\left\langle u_2^* u_4\right\rangle + \left\langle u_1^* u_4\right\rangle\left\langle u_2^* u_3\right\rangle$ to Eq. 8, a relation between the second order and first order correlation function is obtained,

$$G^{(2)}\left(x_1',x_2',x_1'',x_2''\right)=G^{(1)}\left(x_1',x_1''\right)G^{(1)}\left(x_2',x_2''\right)$$
$$+G^{(1)}\left(x_1',x_2''\right)G^{(1)}\left(x_2',x_1''\right), \tag{10}$$

where $G^{(1)}\left(x_i,x_j\right)\equiv\left\langle\psi^*\left(x_i\right)\psi\left(x_j\right)\right\rangle$. Substitution of Eq. 10 into Eq. 9 yields

$$\begin{aligned}
&G^{(2)}\left(x_1,x_2\right)\\
&=\left(\int dx_1'dx_1''G^{(1)}\left(x_1',x_1''\right)h_1\left(x_1,x_1'\right)h_1^*\left(x_1,x_1''\right)\right)\\
&\times\left(\int dx_2'dx_2''G^{(1)}\left(x_2',x_2''\right)h_2\left(x_2,x_2'\right)h_2^*\left(x_2,x_2''\right)\right)\\
&+\left(\int dx_1'dx_2''G^{(1)}\left(x_1',x_2''\right)h_1\left(x_1,x_1'\right)h_2^*\left(x_2,x_2''\right)\right)\\
&\times\left(\int dx_2'dx_1''G^{(1)}\left(\mathrm{x}_2',\mathrm{x}_1''\right)h_2\left(x_2,x_2'\right)h_1^*\left(x_1,x_1''\right)\right)\\
&=\left\langle I_1\left(x_1\right)\right\rangle\left\langle I_2\left(x_2\right)\right\rangle\\
&+\left|\int dx_1'dx_2'h_1\left(x_{1,}x_1'\right)h_2^*\left(x_{2,}x_2'\right)G^{(1)}\left(x_1',x_2'\right)\right|^2.
\end{aligned} \tag{11}$$

Therefore, the deviation of the second order correlation function is

$$\begin{aligned}
&\Delta G^{(2)}\left(x_1,x_2\right)\\
&\equiv G^{(2)}\left(x_1,x_2\right)-\left\langle I_1\left(x_1\right)\right\rangle\left\langle I_2\left(x_2\right)\right\rangle\\
&\equiv\left|\int dx_1'dx_2'h_1^*\left(x_{1,}x_1'\right)h_2\left(x_{2,}x_2'\right)G^{(1)}\left(x_1',x_2'\right)\right|^2.
\end{aligned} \tag{12}$$

For $z\gg x-x'$ the approximation $l=z+\dfrac{\left(x-x'\right)}{2z}$ is made and applied to Eq. 2. This leads to

$$h_1\left(x,x'\right)=h_2\left(x,x'\right)=e^{ikz}e^{i\frac{k\left(x-x'\right)^2}{2z}}T\left(x'\right), \tag{13}$$

together with

$$G^{(1)}\left(x_1',x_2'\right)=I_0\delta\left(x_1'-x_2'\right). \tag{14}$$

Up to this point, no distinction was made between the dephaser or decoherer. For the dephaser, substitution of Eq. 13 and 14 into Eq. 12, leads to the deviation,

$$\begin{aligned}&\Delta G^{(2)}\left(x_1,x_2\right)\\&\equiv\left|\int dx_1'dx_2'h_1^*\left(x_{1,}x_1'\right)h_2\left(x_{2,}x_2'\right)G^{(1)}\left(x_1',x_2'\right)\right|^2\\&=\left|2\pi I_0e^{ik\frac{\left(x_2^2-x_1^2\right)}{2z}}\widetilde{|T|^2}\left(\frac{k}{z}\left(x_1-x_2\right)\right)\right|^2.\end{aligned} \tag{15}$$

The $\widetilde{|T|^2}$ is the Fourier transformation of $|T|^2$. We recall the expression of the transmission function (Eq. 3) and substitute in Eq. 15. The dephaser will be cancelled so that only the double-slit transmission function remains:

$$\widetilde{T^2}\left(\frac{k}{z}\left(x_1-x_2\right)\right)=\widetilde{D^2}\left(\frac{k}{z}\left(x_1-x_2\right)\right). \tag{16}$$

Thus, Eq. 15 can be rewritten as

$$\begin{aligned}&\Delta G^{(2)}\left(x_1,x_2\right)\\&=\left|2\pi I_0e^{ik\frac{\left(x_2^2-x_1^2\right)}{2z}}\widetilde{D^2}\left(\frac{k}{z}\left(x_1-x_2\right)\right)\right|^2.\end{aligned} \tag{17}$$

Using the normalized correlation function:

$$\begin{aligned}&\Delta g^{(2)}\left(x_1,x_2\right)\\&=\Delta G^{(2)}\left(x_1,x_2\right)/\left(\left\langle I_1\left(x_1\right)\right\rangle\left\langle I_2\left(x_2\right)\right\rangle\right)\\&=\left\langle I_1\left(x_1\right)I_2\left(x_2\right)\right\rangle/\left(\left\langle I_1\left(x_1\right)\right\rangle\left\langle I_2\left(x_2\right)\right\rangle\right)-1,\end{aligned} \tag{18}$$

we come to a result that

$$\Delta g^{(2)}\left(x_1,x_2\right)\propto\left|e^{ik\frac{\left(x_2^2-x_1^2\right)}{2z}}\widetilde{D^2}\left(k_{x_1,x_2}\right)\right|^2, \tag{19}$$

where $k_{x_1,x_2}=k\left(x_1-x_2\right)/z$. In the simulation symmetric coordinates are chosen in the x-axis, $x=x_1=-x_2$ and $x=0$ is chosen at the center of detection screen. The normalized second order correlation function [5] is thus

$$\Delta g^{(2)}\left(x,-x\right)\equiv\Delta g^{(2)}\left(x\right)\propto\left|\widetilde{D^2}\left(2kx/z\right)\right|^2. \tag{20}$$

This result states that the normalized correlation function is proportional to the Fourier transformation of the double-slit spatial pattern and thus the far field interference pattern. It reveals that the diffraction pattern will be recovered if a dephaser is applied. However, for a decoherer, the far-field pattern is not recovered. For the decoherer, Eq. 15 is expressed as

$$\begin{aligned}&\Delta G^{(2)}\left(x_1,x_2\right)\\&\equiv\sum_n\left|\int dx_1'dx_2'h_1^*\left(x_{1,}x_1'\right)h_2\left(x_{2,}x_2'\right)G_n^{(1)}\left(x_1',x_2'\right)\right|^2\\&=\sum_n\left|2\pi I_0 e^{ik\frac{\left(x_2^2-x_1^2\right)}{2z}}\widetilde{T_n^2}\left(\frac{k}{z}\left(x_1-x_2\right)\right)\right|^2,\end{aligned} \tag{21}$$

where $G_n^{(1)}\left(x_i',x_j'\right)$ is the $n$-th wave's correlation function.

The deviation of the normalized correlation function of the far field intensity distribution for 500 different phase realizations are averaged using Eq. 17 and compared to the Fourier transformation of the double-slit transmission function,

$$\begin{aligned}&I=4d^2\left(\frac{d\sin\theta}{\lambda}\right)^{-2}\operatorname{sinc}^2\left(\frac{d\sin\theta}{\lambda}\right)\times\\&\cos^2\left(\frac{\pi D\sin\theta}{\lambda}\right)\end{aligned}, \tag{22}$$

where $d$ is the width of slits, $D$ is the distance between the center of two slits and $\lambda$ is wavelength (30 pm). The angle $\theta$ is related to detector position by $\theta=\frac{2x}{L_2}$, where $x$ is the distance to the center of the detection screen, and $L_2$ is the distance between detection screen and double-slit. The result is shown in Fig. 6.

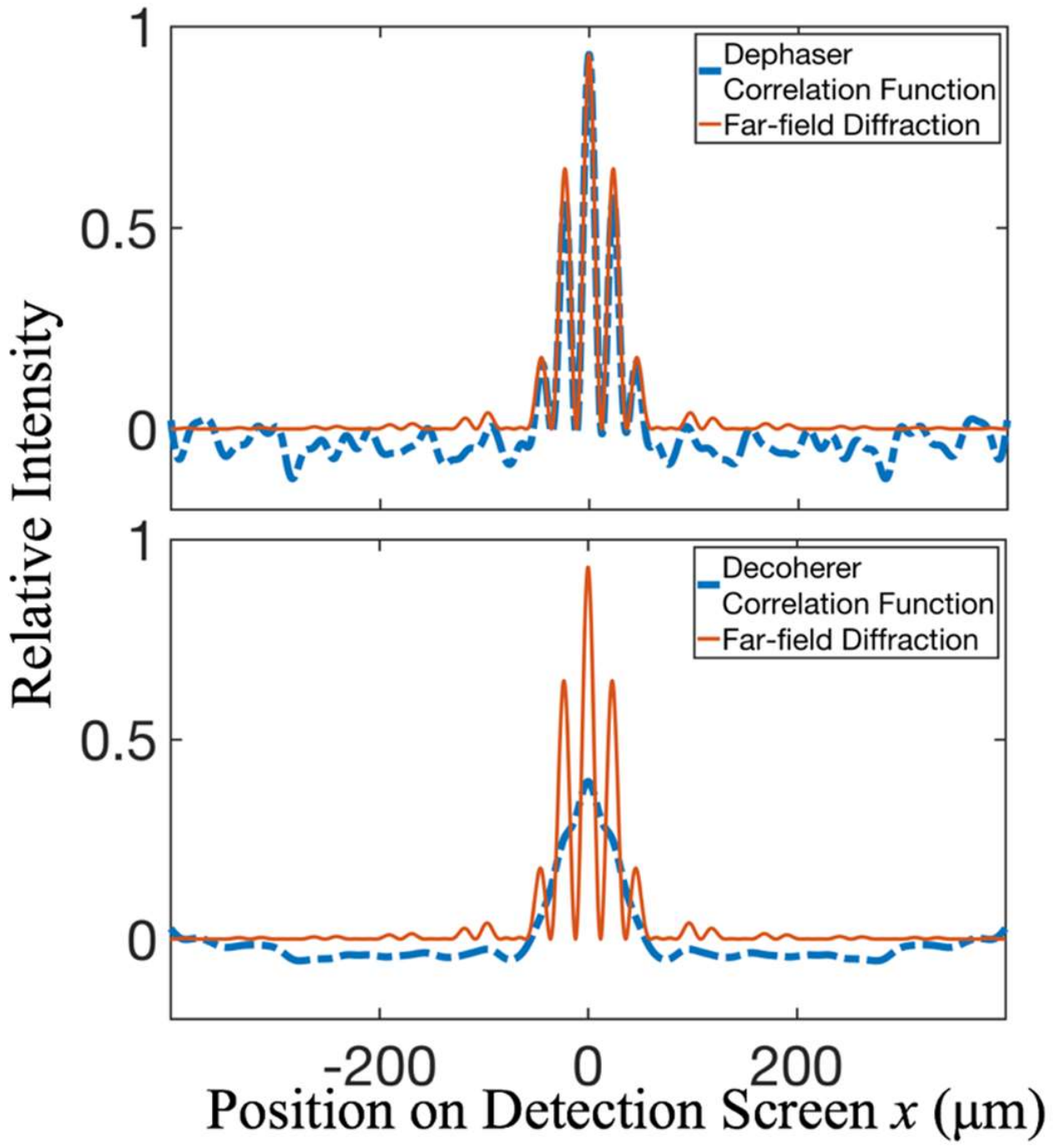

FIG. 6. Double slit pattern recovery. A comparison is made between the deviation of second order correlation function and the corresponding far-field diffraction pattern (Eq. 22) of the double-slit. The orange solid curve is the far-field double slit diffraction pattern. In (a), the blue dashed curve is the correlation function averaged over 500 dephaser realizations, while in (b) the blue line is the same for 500 decoherer realizations. In (a), the deviation of second order correlation function recovers the diffraction pattern while in (b) this is not the case.

The relationship between Fig.5 and 6 is that two example patterns of the 500 total simulations are shown for the dephaser in Fig. 5(a) and Fig. 5(c), which correspond to the correlation function in 6a. In the same way the decoherer patterns in Fig. 5(b) and Fig. 5(d), correspond to Fig. 6(b). The effect of applying the correlation function on the patterns that underwent dephasing (blue in Fig 6(a)) is in agreement with the theoretical double slit diffraction pattern (orange) which confirms Eq. 17. This phenomenon has been observed in the optical regime [5] and we have now shown that it is possible to be observed in the matter-wave regime. In Fig. 6(b) the correlation function for the decoherer shows the absence of a double-slit pattern.

In summary, the increase of entropy (Fig. 3) is one-to-one related to the absence of a double-slit diffraction pattern in the correlation function (Fig. 6), and such an absence identifies the presence of a time-irreversible process. This finding regarding the change in entropy matches Zurek's description of decoherence very well. Zurek [23] explains that "*Observers can be ignorant of phases for reasons that do not lead to an imprint of the state of the system on the environment. …Transfer of information about a decohering system to the environment is essential, and plays a key role in the interpretation.*" That is to say, if information is transferred by decoherence to the environment, then the information entropy will increase. *"Hence, in the case of dephasing …, information about the cause obtained …afterwards,* *suffices to undo the effect."* In our example, the effect of the dephasing (the scrambling of the diffraction image) is undone by the correlation method presented here. *"Decoherence relies on entangling interactions… Thus neither prior nor posterior knowledge of the state of the environment is enough."* Indeed, for our example, the correlation method does not recover the diffraction image for the case of decoherence.

It is not clear at this point how general the correlation function processing method is. When the dephaser only acts at the location of one slit, and the dephaser is turned off at the location of the other slit, the double-slit pattern is not recovered in the correlation function. This is an example where the method fails. A more detailed study is required to identify both analytically and numerically what the validity range of the correlation method is. In this context the interesting approach of Stibor [24–26] to remove dephasing is relevant. In their method correlation in space and time is used and experimentally shown to remove dephasing for externally applied fields. It is different from the present method in that a specific form of the dephasing fields is assumed, that is uses time explicitly, and that it does not evaluate decoherence.

All our numerical tests indicate that our method works when the dephaser and decoherer pervade the entire wavefunction, with spatial fluctuations one order of magnitude (or more) smaller than the single slit width. This is the scenario discussed in multiple theoretical models [14,16,17,27,28] and experiments [10,13,18]. An additional experimental requirement is that multiple probabilities distributions have to be recorded, each corresponding to different phase realizations. This can be done as a function of the position of the decohering or dephasing object. For example, the lateral position of the surface in Fig. 2 could be varied. For a position dependent dephaser, it is expected that the diffraction pattern will be recovered, while for a decoherer (as for example based on image charge) the lateral position does not affect the diffraction image and the diffraction pattern will not be recovered. The correlation method is not expected to work in all cases. For example, if the surface acts as a homogeneous decoherer, there may be no position dependence. If the time dependence of the interaction between the electron and the surface is on the order of the electron-electron interaction time ($\approx 10$ fs) then a repetitive time-averaged accumulation of the probability distribution will yield identical results. In this case techniques exists that are developed for ultrafast electron diffraction and microscopy [29,30] where this time domain in now being reached [31].

In conclusion, an alternative to holography or tomography is offered to distinguish dephasing from decoherence and thus identify time reversible from time irreversible processes. This technique of using repetitive correlation measurements to distinguish between dephasing and decoherence is discussed in the context of electron matter optics, should be applicable to optics, and can lead to new experiments in both fields.

The authors wish to thank Wolfgang Schleich, Sam Keramati and Eric Jones for discussions. This work was completed utilizing the Holland Computing Center of the University of Nebraska, which receives support from the Nebraska Research Initiative. We gratefully acknowledge support by the U.S. National Science Foundation under Grant No. 1602755.

[1] M. A. Nielsen and I. L. Chuang, *Quantum Computation and Quantum Information* (Cambridge University Press, 2011).
[2] G. Mauro D'Ariano, M. G. A. Paris, and M. F. Sacchi, in *Adv. Imaging Electron Phys.*, edited by P. W. Hawkes (Elsevier, 2003), pp. 205–308.
[3] R. P. Feynman, R. B. Leighton, and M. Sands, in *Feynman Lect. Phys.* (Addison-Wesley, Reading, MA, 1965), p. Chapter 1.
[4] R. Bach, D. Pope, S.-H. Liou, and H. Batelaan, New J. Phys. **15**, 033018 (2013).
[5] L. Rui-Feng, Y. Xin-Xing, F. Yi-Zhen, Z. Pei, Z. Yu, G. Hong, and L. Fu-Li, Chin. Phys. B **23**, 054202 (2014).
[6] M. Brune, E. Hagley, J. Dreyer, X. Maître, A. Maali, C. Wunderlich, J. M. Raimond, and S. Haroche, Phys. Rev. Lett. **77**, 4887 (1996).
[7] C. J. Myatt, B. E. King, Q. A. Turchette, C. A. Sackett, D. Kielpinski, C. A. Sackett, C. J. Myatt, D. J. Wineland, C. Monroe, B. E. King, W. M. Itano, Q. A. Turchette, W. M. Itano, C. Monroe, and D. J. Wineland, Nature **403**, 269 (2000).
[8] K. Hornberger, S. Uttenthaler, B. Brezger, L. Hackermüller, M. Arndt, and A. Zeilinger, Phys. Rev. Lett. **90**, 160401 (2003).
[9] L. Hackermüller, K. Hornberger, B. Brezger, A. Zeilinger, and M. Arndt, Nature **427**, 711 (2004).
[10] P. Sonnentag and F. Hasselbach, Phys. Rev. Lett. **98**, 200402 (2007).
[11] J. Trost and K. Hornberger, Phys. Rev. Lett. **103**, 023202 (2009).
[12] E. R. Jones, R. A. Bach, and H. Batelaan, Eur. J. Phys. **36**, 065048 (2015).
[13] B. Barwick, G. Gronniger, L. Yuan, S.-H. Liou, and H. Batelaan, J. Appl. Phys. **100**, 074322 (2006).
[14] J. R. Anglin and W. H. Zurek, in *Dark Matter Cosmol. Quantum Meas. Exp. Gravit.* (Editions Fronti'eres, Gif-sur-Yvette, France, Les Arcs, Savoie, France, 1996), pp. 263–270.
[15] J. R. Anglin, J. P. Paz, and W. H. Zurek, Phys. Rev. A **55**, 4041 (1997).
[16] P. Machnikowski, Phys. Rev. B **73**, 155109 (2006).
[17] S. Scheel and S. Y. Buhmann, Phys. Rev. A **85**, 030101 (2012).
[18] F. Röder and A. Lubk, Ultramicroscopy **146**, 103 (2014).
[19] R. P. Feynman, Rev. Mod. Phys. **20**, 367 (1948).
[20] J. Cheng and S. Han, Phys. Rev. Lett. **92**, 093903 (2004).
[21] P. Kazemi, S. Chaturvedi, I. Marzoli, R. F. O'Connell, and W. P. Schleich, New J. Phys. **15**, 013052 (2013).
[22] J. W. Goodman, in *Stat. Opt.* (Wiley, 2000), pp. 41–44.
[23] W. H. Zurek, Rev. Mod. Phys. **75**, 715 (2003).
[24] A. Rembold, G. Schütz, W. T. Chang, A. Stefanov, A. Pooch, I. S. Hwang, A. Günther, and A. Stibor, Phys. Rev. A **89**, 033635 (2014).
[25] A. Günther, A. Rembold, G. Schütz, and A. Stibor, Phys. Rev. A **92**, 053607 (2015).
[26] A. Rembold, G. Schütz, R. Röpke, W. T. Chang, I. S. Hwang, A. Günther, and A. Stibor, New J. Phys. **19**, 033009 (2017).
[27] A. Howie, Ultramicroscopy **111**, 761 (2011).

[28] A. Howie, J. Phys. Conf. Ser. **522**, 012001 (2014).
[29] B. J. Siwick, J. R. Dwyer, R. E. Jordan, and R. J. D. Miller, Science **302**, 1382 (2003).
[30] A. H. Zewail, Annu. Rev. Phys. Chem. **57**, 65 (2006).
[31] M. Gulde, S. Schweda, G. Storeck, M. Maiti, H. K. Yu, A. M. Wodtke, S. Schäfer, and C. Ropers, Science **345**, 200 (2014).